# The risks of risk assessment: causal blind spots when using prediction models for treatment decisions

Nan van Geloven, PhD (1*), Ruth H Keogh, PhD (2), Wouter van Amsterdam, MD, PhD (3), Giovanni Cinà, PhD (4,5,6), Jesse H Krijthe, PhD (7), Niels Peek, PhD (8,9), Kim Luijken, PhD (3), Sara Magliacane, PhD (5), Paweł Morzywołek, PhD (10,11), Thijs van Ommen, PhD (12), Hein Putter, PhD (1), Matthew Sperrin, PhD (8), Junfeng Wang, PhD (12), Daniala L Weir, PhD (12), Vanessa Didelez, PhD (13,14)

(1) Leiden University Medical Center, Leiden, The Netherlands
(2) London School of Hygiene and Tropical Medicine, London, United Kingdom
(3) University Medical Center Utrecht, Utrecht, The Netherlands
(4) Amsterdam University Medical Centers, Amsterdam, The Netherlands
(5) University of Amsterdam, Amsterdam, The Netherlands.
(6) Pacmed, Amsterdam, The Netherlands
(7) Delft University of Technology, Delft, The Netherlands
(8) University of Manchester, Manchester, United Kingdom
(9) THIS Institute, University of Cambridge, United Kingdom
(10) Ghent University, Ghent, Belgium
(11) University of Washington, Seattle, United States
(12) Utrecht University, Utrecht, The Netherlands
(13) Leibniz Institute for Prevention Research and Epidemiology - BIPS, Bremen, Germany
(14) Department of Mathematics and Computer Science, University of Bremen, Germany

*corresponding author: n.van_geloven@lumc.nl

**Short title**: The risks of risk assessment

**Word count**: 3320 + figure + table

# Abstract

Clinicians increasingly rely on prediction models to guide treatment choices. Most prediction models, however, are developed using observational data that include some patients who have already received the treatment the prediction model is meant to inform. Special attention to the causal role of those earlier treatments is required when interpreting the resulting predictions.

We identify "causal blind spots" in three common approaches to handling treatment when developing a prediction model: including treatment as a predictor, restricting to individuals taking a certain treatment, and ignoring treatment. Through several real examples, we illustrate how the risks obtained from models developed using such approaches may be misinterpreted and can lead to misinformed decision-making. Our discussion covers issues attributable to confounding, selection, mediation and changes in treatment protocols over time.

We advocate for an extension of guidelines for the development, reporting and evaluation of prediction models to avoid such misinterpretations. Developers must ensure that the intended target population for the model, and the treatment conditions under which predictions hold, are clearly communicated. When prediction models are intended to inform treatment decisions, they need to provide estimates of risk under the specific treatment (or intervention) options being considered, known as 'prediction under interventions'. Next to suitable data, this requires causal reasoning and causal inference techniques during model development and evaluation. Being clear about what a given prediction model can and cannot be used for prevents misinformed treatment decisions and thereby prevents potential harm to patients.

# Introduction

Clinical prediction (or prognostic) models estimate the risk of future outcomes based on observable patient characteristics. They do so by learning from a historical, often observational, dataset (the development dataset) in which outcomes are already known. Partly fuelled by the recent surge in artificial intelligence, clinicians increasingly use these models, expecting that knowing an individual's risk will help them make better decisions on medical treatments or lifestyle advice that may lower that risk (e.g., (1)). For example, clinicians may choose to treat high risk patients more aggressively, while managing low risk patients more conservatively. In this paper, we illustrate how using clinical prediction models derived from observational data in this way may inadvertently lead to inappropriate or even harmful treatment decisions.

We focus on the use of predictions to inform a decision on a particular treatment (the 'target treatment'), but the problems we point out apply equally to, say, lifestyle advice. The key issue is that most prediction models are derived from observational data in which some individuals have already received the target treatment.(2) Predictions will be influenced by the mix of patients who did or did not receive these target treatments in the development data, as well as the reasons (indications) for those treatment choices. This makes predictions challenging to interpret: for instance, a new patient may have a low estimated risk because similar patients in the development data received aggressive treatments that were successful, hence conservative management is likely inappropriate.

A patient and doctor deciding on a target treatment would ideally know the patient's risk 'if they were to take the treatment' and their risk 'if they were not to take it'. We will refer to such risks that carry a 'what-if' interpretation as potential risks. A difference between the potential risks with and without treatment indicates a causal treatment effect. While treatment decisions may be based on the size of the causal treatment effect, the two separate potential risks under treatment and under no treatment provide more complete information. In certain cases, knowing the patient's potential risk under a single treatment option could suffice for the treatment decision. For instance, if a patient's risk is low if they were to remain untreated, treatment may be deemed unnecessary, irrespective of the size of the causal treatment effect. Alternatively, if a patient's prognosis is poor even if they were to be treated, this may prompt the decision to refrain from treatment.

As we will illustrate in this paper, many prediction models derived from observational data using standard methods do not provide valid estimates of causal treatment effects, nor do they estimate the potential risk under a given treatment option properly. The methodology for developing models that can do this properly has been called "counterfactual prediction" or "prediction under interventions" (3,4,5). These methods link causal inference approaches to prediction methods, which traditionally have been treated separately.

In this paper, we discuss the risks of falsely assigning a 'what-if' interpretation to prediction outputs. We attribute these misinterpretations to 'causal blind spots' and explain how they may arise under three common ways of handling treatments when developing (or training) a prediction model: 1) including the target treatment as a predictor, 2) restricting the development data to those who did (or those who did not) receive the target treatment, and 3) ignoring the target treatment. We then outline recommendations for deriving and evaluating algorithms that do allow for a 'what if' interpretation.

## Causal blind spots when including treatment as a predictor

Suppose that a prediction model was developed on observational data and included the baseline target treatment status as one of the predictors. Patients and doctors who want to know the patient's potential risk if they were to take and if they were not to take the target treatment, might be inclined to specify 'yes' respectively 'no' for the treatment variable and interpret the resulting predictions as such. Below we describe three situations in which this interpretation is not valid, illustrated by real examples. The assumed causal structure linking the relevant variables underlying these examples are depicted using causal diagrams (directed acyclic graphs-DAGs) in Figure 1.

### *Confounding – blood pressure lowering medication and cardiovascular risk*

It is well known that associations found in prediction models developed using observational data cannot be interpreted causally due to confounding, i.e., predictors and outcome may have common causes, not included in the model, that induce their apparent relationship.(6) What is less appreciated is that, for the same reason, risks obtained from these models cannot be interpreted as representing potential risks under specified treatment options. Consider the situation where an individual wants to know their potential cardiovascular risk if they were or were not to use blood pressure lowering medication. The PREDICT model which estimates 5-year risk of cardiovascular disease and was derived from primary care data from over 400,000 individuals in New Zealand, includes as a predictor an indicator of use of blood pressure lowering medication at baseline (prevalent and incident use). Use of this medication is associated with a *higher* risk of developing cardiovascular disease in the model.(7) Applying the prediction model for a 75 year old woman who is on blood pressure lowering medication, smokes, has systolic BP 120 and TC/HDL ratio 3.5, results in an estimated 5-year risk of cardiovascular disease of 15%. Using the same inputs, except switching to no use of blood pressure lowering medication, gives an estimated risk of only 11%. This should certainly not be interpreted as evidence in support of discontinuing the blood pressure lowering medication for this woman. Instead, the higher risk could be explained by pre-existing health conditions in people receiving anti-hypertensive medication which are also related to the outcome, but which are not included as predictors. For example, a prior diagnosis of hypertension, which is not included in the model, may have prompted the initiation of medication while simultaneously increasing cardiovascular risk. This is sometimes referred to as 'confounding by indication' (Figure 1A).

### *Selection – BMI and mortality risk*

Interpretations of predictions are also impacted by how patients are selected for inclusion in observational development data, in particular when they are selected based on a particular feature that is causally related to both the target treatment and the outcome.(8,9,10) Consider, for example, the question whether a patient with a heart condition should or should not gain weight. For such lifestyle advice, doctors would want to know a patient's potential risk if they were to gain weight or were not to do so. Several prediction models developed in datasets selecting exclusively patients with a heart condition have included baseline BMI as a predictor (which as a modifiable lifestyle factor is the target treatment in this example) and reported, somewhat unexpectedly, that patients with obesity (high BMI) had lower mortality risk compared to patients without obesity (e.g., (11,12)). This should certainly not lead to advising heart patients to gain weight to improve their survival. Instead, the underlying reason for the apparently paradoxical protective survival effect of obesity could be that obesity is a major cause of heart problems and patients in the development data *without* obesity were more likely to have other (unmeasured) factors (e.g., genes) that led to their heart condition. If these characteristics are also risk factors for mortality, then, in persons with a heart condition, obesity may act as a marker for the *absence* of much more serious risk factors.(13, Figure 1B) Lifestyle interventions to increase BMI will not eliminate these other risk factors. Similar spurious protective effects of obesity have been reported in other settings where patients were selected based on underlying health conditions such as diabetes and this has been referred to as the 'obesity paradox'.(13,14)

In the supplementary material we give two additional examples of prediction models where the patient selection likely introduced bias.

### *Mediation – smoking and risk of covid-19 death*

A third situation where a prediction model does not provide risks that can inform on alternative treatment options, is when the model includes a consequence of the target treatment as predictor. Williamson et al. evaluated risk factors associated with covid-19-related death using a multivariable model fitted using primary care records of >17 million adults in the United Kingdom.(15) The multivariable model suggested a lower hazard of covid-19-related death for smokers compared with non-smokers. Although the authors warned against causal interpretation of their model and did not advise its use for decision making, policy-makers initially interpreted this result as evidence that smoking was protective for covid-19-related death, which nearly resulted in smokers being excluded from shielding policies in France.(16,17)

One likely mechanism contributing to the counterintuitive finding regarding smoking is that the model included factors that are causal consequences of smoking, such as respiratory diseases. If these factors are also causes of the outcome, they may have acted as mediators, i.e., intermediary variables transmitting the influence of smoking to mortality (Figure 1C). Including mediators in the model may mask (or 'adjust away') part of the true causal effect of smoking. It might be assumed that mediators cannot be included in a prediction model, given that predictors are typically measured at the same baseline moment, and there is no time for the treatment variable to impact the mediator. However, the 'treatment variable' (here, smoking) often represents a prevalent exposure, which may have influenced other factors included as predictors (here, development of respiratory diseases). In Supplementary material B we point out a second mechanism that may be at play in this example.

## Causal blind spots when restricting the development data based on the target treatment

Another approach that is often taken in an attempt to obtain potential risks *if a patient were (or were not) to take treatment*, is to restrict the observational development data to patients who did (or did not) use the target treatment.(18) However, except in historical cohorts where treatment was not yet available, individuals who are (un)treated in observational data are generally not representative of those for whom that decision is still to be made. In fact, restricting the development data based on treatment status brings the same issues as the earlier examples in which treatment was included as a predictor in the model, as well as potentially introducing issues due to selection similar to those described in the above BMI example where selection was not on the basis of treatment.

For instance, in the above PREDICT model, suppose that in an attempt to estimate patients' potential risks if they were not to use blood pressure lowering medication, investigators had restricted the development data to individuals not using blood pressure lowering medication at baseline. Applying this model to the example patient considered earlier (75-year old woman, smokes, systolic BP 120, TC/HDL ratio 3.5) to obtain a 5-year risk of cardiovascular disease under the condition of not using blood pressure lowering medication would again output a risk of around 11%. This risk is likely based on relatively healthy individuals (e.g., without prior diagnosis of hypertension) and, as before, it is not a valid estimate of the potential risk if the example patient were to stop taking the treatment. Additionally, immortal time bias can occur if the development data are restricted to untreated individuals who do not initiate blood pressure-lowering medication during follow-up, as this restriction would be based on post-baseline treatment status. (19,20)

Similarly, prediction models developed on observational data from patients who have all received the target treatment will only be applicable to patients who would receive the treatment under the treatment assignment strategy used in the development data. Such models may not be valid for a wider group of patients for whom the decision of whether to treat is yet to be made, as patients who did not receive treatment under the earlier assignment strategy may differ from those who did in ways not captured by the prediction model. It is not uncommon for prediction models to be developed using data restricted to patient receiving the target treatment. A systematic review of prognostic models for outcomes of extracorporeal membrane oxygenation (ECMO) therapy found that all 58 included observational studies restricted to patients receiving the ECMO therapy. Meanwhile, 11 of these 58 studies explicitly stated that their primary aim was to inform decisions regarding ECMO initiation for new patients.(21) Another example is the EuroSCORE model, which predicts mortality after cardiac surgery. It was suggested the model could be used to assist in deciding whether to proceed with the surgery based on the estimated risk, but the model was developed on an observational cohort consisting solely of patients who had undergone the surgery.(22)

## Causal blind spots when ignoring treatments applied in the development data

Given the complications outlined above, one may think it is better to ignore treatment status when developing a prediction model – that is, not to include the target treatment as predictor, nor to restrict based on treatment status. When the development data contains a mix of patients – some treated and some untreated – ignoring treatment in the prediction model will result in risks that are only valid under continuation of the implicit treatment policy in the observational development data, which we refer to as risks 'under current care'.

Now the question arises as to whether risks 'under current care' are suitable for supporting treatment decisions in new patients. One may argue that if the risk under current care is high, the patient should receive more aggressive treatment. But more aggressive treatment will be different from current care and since we do not have a prediction under this novel treatment strategy, we don't know if and how much it would benefit the patient. Moreover, what constitutes 'current care' is often poorly defined, making it difficult to determine what it would mean for a particular individual and hence what would constitute more aggressive treatment options. Conversely, one might be tempted to conclude that those with low risk under current care could receive less intense treatment. This advice is problematic because a patient may have low risk precisely because they received appropriate treatment under current care. Withholding treatment from such 'low risk' patients may lead to harm.(23) The potential danger of this approach was highlighted in the context of a prediction model estimating mortality risk among patients hospitalized with pneumonia.(24) The goal was to identify pneumonia patients with a low mortality risk who could be safely treated as outpatients while those at high risk could be admitted to hospital. This implicitly assumes that a pneumonia patient treated in the hospital with a low mortality risk would have similarly experienced low risk if treated at home. This assumption has later been called into question because the prediction algorithm estimated that patients with asthma had a low mortality risk. However, this was likely because these patients received effective intensive hospital care in the development data, leading to favourable outcomes.(25) If new asthma patients were instead treated as outpatients based on the low risk under current care, their outcomes would likely be worse than predicted, and crucially, worse than under the current care before introducing the prediction model.

Similar issues have been highlighted for more recently-developed prediction models such as QRISK3, which estimates the 10-year risk of cardiovascular disease in the general population and is used in primary care in the UK for supporting decisions on initiating statins.(26) Ideally those decisions would be supported by estimates of potential risks if patients were not to use statins; if this untreated risk is low, no preventive treatment would be needed. During model development however, initiation of statin therapy (as well as other treatments) during follow-up was ignored. This means that the risks obtained from the model represent expected outcomes under the statin initiation policy in the development data. An individual might have a low risk according to this model because similar patients initiated treatments at a later stage, and therefore, they would not necessarily have a low risk without treatment.(27,28)

Both examples illustrate the *prediction paradox*: when predictions are used to support treatment decisions, this will change treatment practice and thus potentially invalidate any prediction that was made under historical practice.(29,30,31) For this reason, ignoring treatments during model development in general does not provide risks that can be used to guide future treatment decisions.

## The way forward

The notion that "prognosis cannot be divorced from contemplated medical action, nor from action to be taken by the patient in response to prognostication" was recognized as early as 1987 (32,33), however, using established and implemented prediction models as examples, we demonstrated that misinterpretations of prediction models remain widespread.

As a way forward, we argue that developers of clinical prediction models should more carefully describe the target treatment(s) the prediction model is intended to inform, how that treatment was assigned in the development data, and how it was handled during model development.(34) An effective way of specifying the role of the target treatment in a prediction model is by using the so-called *prediction estimand* framework.(27,35,36) In addition to the usual elements required in reporting of prediction models (outcome, target population, moment of intended use, predictors etc), the prediction estimand formally describes the treatment conditions under which predictions hold (e.g. risk under current care, and how such care is defined). For decision support, the prediction estimand should specify the level(s) of the target treatment for which the model provides predictions (e.g. the potential risk 'if the individual were not to start the target treatment').(35) The prediction estimand should be part of the planning stage of model development and evaluation, and, importantly, communicated to end users of the model.(36,37) To encourage its use, a description of the prediction estimand could be added to existing reporting guidelines such as TRIPOD+AI and appraisal tools such as PROBAST+AI.(34,38,39,40)

Recent literature delineates how causal inference techniques for estimating treatment effects when using observational data can be applied (or modified) for valid development and evaluation of potential risks under different treatment options.(3,5,27,35,41-45) Our examples highlight that careful consideration is needed around selection of the data used for model development, and of the variables included as predictors. As well as collecting predictors of the outcome, additional variables required to control for confounding (which may be time-varying) and to avoid selection bias are needed, plus information on starting and stopping of treatment.(4) To avoid time related bias, the data source must allow for alignment of eligibility, treatment assignment and start of follow at the time of decision making.(46) The assumptions allowing valid prediction under interventions using observational data need to be carefully stated, justified and evaluated. For a summary of recommendations for prediction under interventions, see Table 1.

Randomised controlled trials are the only setting where purely predictive methods may produce potential risks under different treatment options, as randomisation ensures that, on average, treatment arms are comparable at baseline. Trial data have been used for this purpose, especially in settings with heterogeneity in treatment effects across individuals.(47,48) However, trial data may not be suitable for widely applicable individualized prediction due to strict inclusion criteria and small sample sizes. While trials allow estimation of potential risks under baseline treatment options, they may be insufficient for potential risks under longer periods of treatment exposure (i.e., sustained treatments) when there is imperfect treatment adherence or drop-out. In that case, again, causal inference methods for prediction under sustained interventions are required.(49) Combining trial data with observational data has been proposed to extend trial results to a broader target population.(50)

# Conclusion

We have identified ‘causal blind spots’ that make prediction models derived from observational data susceptible to misinterpretation and often unsuitable for informing individual treatment decisions.

There is a deluge of prediction models in medical journals, on websites and in apps. With the growing availability of mostly observational data and the rising popularity of machine learning techniques, the future will likely see an increase in black-box algorithms proposed as decision support tools. While flexible methods and large datasets may lead to improved predictive performance, the causal blind spots we have outlined are structural challenges. These problems can only be solved by integrating causal reasoning into predictive modelling.

# Footnotes

**Contributors:** All authors provided a substantial contribution to the design and interpretation of the paper and revised drafts. NvG, RHK and VD initiated the project. NvG wrote the initial draft and is the guarantor for the study. The corresponding author had final responsibility for the decision to submit for publication. The corresponding author attests that all listed authors meet authorship criteria and that no others meeting the criteria have been omitted.

## Funding:

RHK was funded by UK Research and Innovation (Future Leaders Fellowship MR/S017968/1 and MR/X015017/1)

JW was funded by the European Union's Horizon 2020 research and innovation program (The HTx project, grant agreement No 825162)

GC was a consultant for Pacmed during the writing of this article and owns stock appreciation rights (SARs) of the company.

MS acknowledges support of the UKRI AI programme, and the Engineering and Physical Sciences Research Council, for CHAI - Causality in Healthcare AI Hub [grant number EP/Y028856/1]

## Corresponding author:

n.van_geloven@lumc.nl

# References

1. Aleksandrova K, Reichmann R, Kaaks R, et al. Development and validation of a lifestyle-based model for colorectal cancer risk prediction: the LiFeCRC score. BMC Medicine 2021;19:1. Doi:10.1186/s12916-020-01826-0
2. Goodacre S. Using clinical risk models to predict outcomes: what are we predicting and why? Emerg Med J. 2023;40:728–30. doi: 10.1136/emermed-2022-213057
3. Van Geloven N, Swanson SA, Ramspek CL, et al. Prediction meets causal inference: the role of treatment in clinical prediction models. Eur J Epidemiol 2020;35:619–30. Doi:10.1007/s10654-020-00636-1
4. Dickerman BA, Hernán MA. Counterfactual prediction is not only for causal inference. Eur J Epidemiol. 2020;35:615–7. doi: 10.1007/s10654-020-00659-8
5. Lin L, Sperrin M, Jenkins DA, et al. A scoping review of causal methods enabling predictions under hypothetical interventions. Diagn Progn Res. 2021;5:3.
6. Westreich D, Greenland S. The table 2 fallacy: presenting and interpreting confounder and modifier coefficients. Am J Epidemiol. 2013;177:292–8. doi: 10.1093/aje/kws412
7. Pylypchuk R, Wells S, Kerr A, et al. Cardiovascular disease risk prediction equations in 400 000 primary care patients in New Zealand: a derivation and validation study. The Lancet 2018;391:1897–907. doi: 10.1016/S0140-6736(18)30664-0
8. Hernán MA, Hernández-Díaz S, Robins JM. A Structural Approach to Selection Bias: Epidemiology. 2004;15:615–25.
9. Didelez V, Kreiner S, Keiding N. Graphical Models for Inference Under Outcome-Dependent Sampling. Statistical Science. 2010;25:368–87.

10. Hernán MA, Monge S. Selection bias due to conditioning on a collider. BMJ. 2023;381:p1135.
11. Peterson ED, Dai D, DeLong ER, et al. Contemporary Mortality Risk Prediction for Percutaneous Coronary Intervention: Results from 588,398 Procedures in the National Cardiovascular Data Registry. Journal of the American College of Cardiology. 2010;55:1923. doi: 10.1016/j.jacc.2010.02.005
12. Gruberg L, Weissman NJ, Waksman R, et al. The impact of obesity on the short-term and long-term outcomes after percutaneous coronary intervention: the obesity paradox? J Am Coll Cardiol. 2002;39:578–84. doi: 10.1016/s0735-1097(01)01802-2
13. Amundson DE, Djurkovic S, Matwiyoff GN. The obesity paradox. Crit Care Clin. 2010;26:583–96. doi: 10.1016/j.ccc.2010.06.004
14. Lajous M, Banack HR, Kaufman JS, et al. Should Patients with Chronic Disease Be Told to Gain Weight? The Obesity Paradox and Selection Bias. The American Journal of Medicine. 2015;128:334–6. doi: 10.1016/j.amjmed.2014.10.043
15. Williamson EJ, Walker AJ, Bhaskaran K, et al. Factors associated with COVID-19-related death using OpenSAFELY. Nature 2020;584:430–6. doi:10.1038/s41586-020-2521-4
16. Westreich D, Edwards JK, Van Smeden M. Comment on Williamson et al. (OpenSAFELY): The Table 2 Fallacy in a Study of COVID-19 Mortality Risk Factors. Epidemiology 2021;32:e1–2. doi:10.1097/EDE.0000000000001259
17. Tennant P, TABLE 2 FALLACY - or why interpretation needs more than transparency. Presentation at the RSS Interpretable machine learning & causal inference workshop. Dec 20 2015, https://www.youtube.com/watch?v=0S8LZUxi0eg accessed Jan 2nd 2024.
18. Pajouheshnia R, Damen JAAG, Groenwold RHH, et al. Treatment use in prognostic model research: a systematic review of cardiovascular prognostic studies. Diagn Progn Res 2017;1:15. doi:10.1186/s41512-017-0015-0
19. Suissa S, Azoulay L. Metformin and the Risk of Cancer. Diabetes Care. 2012;35:2665–73.
20. Suissa S. Effectiveness of Inhaled Corticosteroids in Chronic Obstructive Pulmonary Disease: Immortal Time Bias in Observational Studies. Am J Respir Crit Care Med. 2003;168:49–53.
21. Pladet LCA, Barten JMM, Vernooij LM, et al. Prognostic models for mortality risk in patients requiring ECMO. Intensive Care Med 2023;49:131–41. doi:10.1007/s00134-022-06947-z
22. Nashef SA, Roques F, Michel P, et al. European system for cardiac operative risk evaluation (EuroSCORE). Eur J Cardiothorac Surg 1999;16:9–13. doi:10.1016/s1010-7940(99)00134-7
23. Van Amsterdam WAC, Van Geloven N, Krijthe JH, et al. When accurate prediction models yield harmful self-fulfilling prophecies. Patterns. 2025;6:101229. doi: 10.1016/j.patter.2025.101229
24. Cooper GF, Aliferis CF, Ambrosino R, et al. An evaluation of machine-learning methods for predicting pneumonia mortality. Artificial Intelligence in Medicine. 1997;9:107–38.
25. Caruana R, Lou Y, Gehrke J, et al. Intelligible Models for HealthCare: Predicting Pneumonia Risk and Hospital 30-day Readmission. In: Proceedings of the 21th ACM SIGKDD International Conference on Knowledge Discovery and Data Mining. Sydney NSW Australia: : ACM 2015. 1721–30. doi:10.1145/2783258.2788613
26. Hippisley-Cox J, Coupland C, Brindle P. Development and validation of QRISK3 risk prediction algorithms to estimate future risk of cardiovascular disease: prospective cohort study. BMJ. 2017;j2099.
27. Sperrin M, Martin GP, Pate A, et al. Using marginal structural models to adjust for treatment drop-in when developing clinical prediction models. Statistics in Medicine. 2018;37:4142–54.
28. Xu Z, Arnold M, Stevens D, et al. Prediction of Cardiovascular Disease Risk Accounting for Future Initiation of Statin Treatment. American Journal of Epidemiology. 2021;190:2000–14.
29. Peek N, Sperrin M, Mamas M, Van Staa T, Buchan I. Hari Seldon, QRISK3, and the prediction paradox. BMJ. 2017;357:j2099.

30. Lenert MC, Matheny ME, Walsh CG. Prognostic models will be victims of their own success, unless…. J Am Med Inform Assoc. 2019;26:1645–50.
31. Liley J, Emerson S, Mateen B, et al. Model updating after interventions paradoxically introduces bias. Proceedings of The 24th International Conference on Artificial Intelligence and Statistics. PMLR 2021:3916–24. https://proceedings.mlr.press/v130/liley21a.html (accessed 5 December 2023)
32. Hilden J, Habbema JDF. Prognosis in medicine: An analysis of its meaning and roles. Theor Med. 1987;8:349–65.
33. Windeler J. Prognosis - what does the clinician associate with this notion? Statistics in Medicine. 2000;19:425–30.
34. Van Amsterdam WAC, Giovanni C, Didelez V, et al. "Prognostic Models for Decision Support Need to Report Their Targeted Treatments and the Expected Changes in Treatment Decisions (Rapid Response)." BMJ, May 30, 2024. https://doi.org/10.1136/bmj-2023-078378/rr-1
35. Luijken K, Morzywołek P, Van Amsterdam W, et al. Risk-based decision making: estimands for sequential prediction under interventions. 2024. Biometrical Journal. 2024;66:e70011
36. Sendak MP, Gao M, Brajer N, et al. Presenting machine learning model information to clinical end users with model facts labels. npj Digit Med. 2020;3:1–4. doi: 10.1038/s41746-020-0253-3
37. Van Royen FS, Asselbergs FW, Alfonso F, et al. Five critical quality criteria for artificial intelligence-based prediction models. European Heart Journal. 2023;44:4831–4.
38. Moons KGM, Damen JAA, Kaul T, et al. PROBAST+AI: an updated quality, risk of bias, and applicability assessment tool for prediction models using regression or artificial intelligence methods. BMJ. 2025;388:e082505. doi: 10.1136/bmj-2024-082505
39. Collins GS, Moons KGM, Dhiman P, et al. TRIPOD+AI statement: updated guidance for reporting clinical prediction models that use regression or machine learning methods. BMJ. 2024;385:e078378.
40. Xu J, Guo Y, Wang F, et al. Protocol for the development of a reporting guideline for causal and counterfactual prediction models in biomedicine. BMJ Open. 2022;12:e059715. doi: 10.1136/bmjopen-2021-059715
41. Van Amsterdam WAC, De Jong PA, Verhoeff JJC, et al. From algorithms to action: improving patient care requires causality. BMC Med Inform Decis Mak. 2024;24:111. https://doi.org/10.1186/s12911-024-02513-3
42. Sperrin M, Diaz-Ordaz K, Pajouheshnia R. Invited Commentary: Treatment Drop-in—Making the Case for Causal Prediction. American Journal of Epidemiology. 2021;190:2015–8.
43. Dickerman BA, Dahabreh IJ, Cantos KV, et al. Predicting counterfactual risks under hypothetical treatment strategies: an application to HIV. Eur J Epidemiol. 2022;37:367–76. doi: 10.1007/s10654-022-00855-8
44. Boyer CB, Dahabreh IJ, Steingrimsson JA. Assessing model performance for counterfactual predictions. 2023. https://doi.org/10.48550/arXiv.2308.13026
45. Keogh RH, Van Geloven N. Prediction under hypothetical interventions: evaluation of counterfactual performance using longitudinal observational data. 2023. https://doi.org/10.48550/arXiv.2304.10005
46. Hernán MA, Sauer BC, Hernández-Díaz S, et al. Specifying a target trial prevents immortal time bias and other self-inflicted injuries in observational analyses. J Clin Epidemiol. 2016;79:70–5. doi: 10.1016/j.jclinepi.2016.04.014
47. Kent DM, Paulus JK, Van Klaveren D, et al. The Predictive Approaches to Treatment effect Heterogeneity (PATH) Statement. Ann Intern Med. 2020;172:35.

48. Hingorani AD, Van der Windt DA, Riley RD, et al. Prognosis research strategy (PROGRESS) 4: Stratified medicine research. BMJ. 2013;346:e5793.
49. Hernán MA, Hernández-Díaz S. Beyond the intention-to-treat in comparative effectiveness research. Clinical Trials. 2012;9:48–55. doi: 10.1177/1740774511420743
50. Colnet B, Mayer I, Chen G, et al. Causal Inference Methods for Combining Randomized Trials and Observational Studies: A Review. Statist Sci. 2024;39. doi: 10.1214/23-STS889

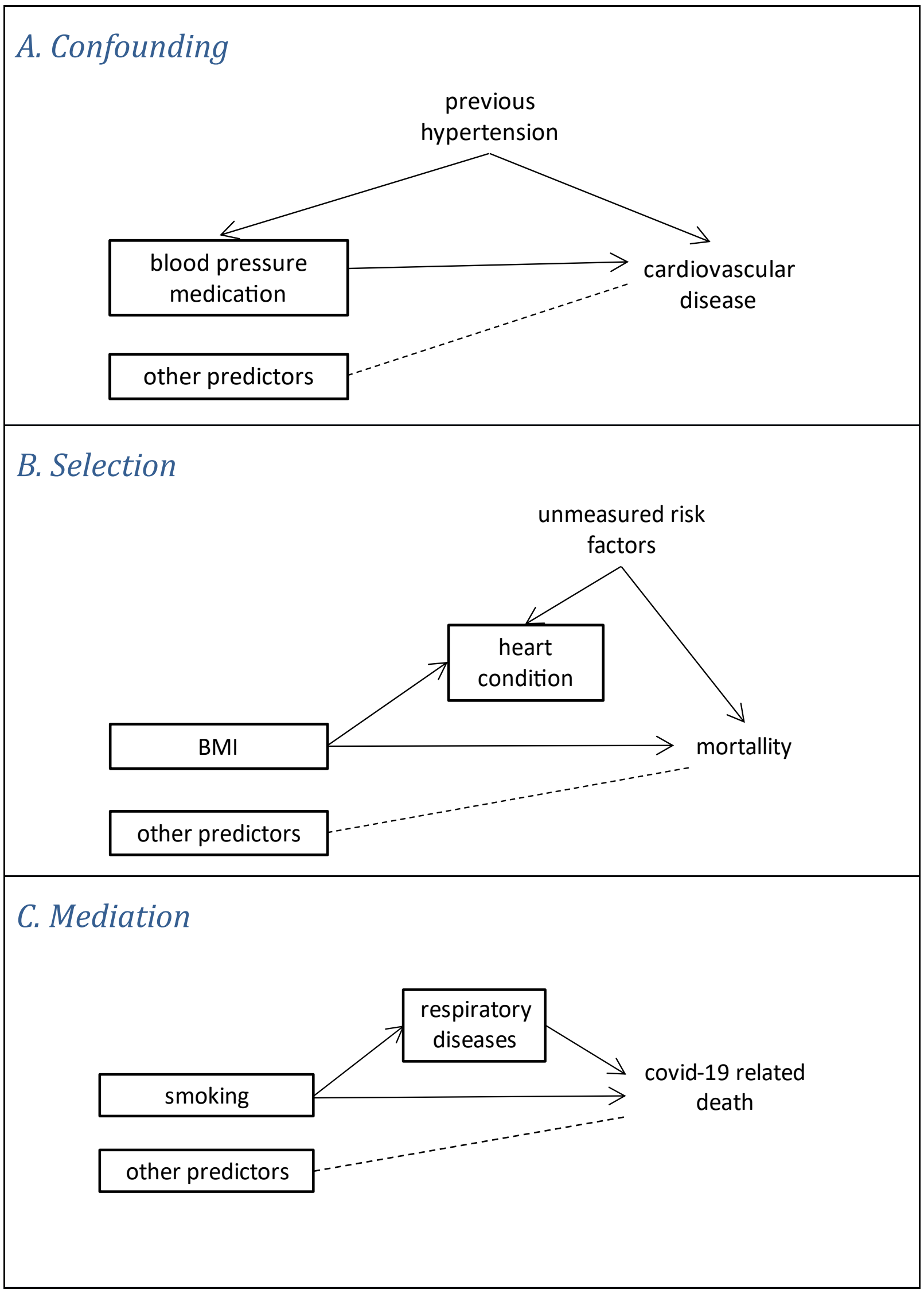

*Figure 1: Causal structures underlying the examples, depicted in causal directed acyclic graphs. A box around a variable indicates that this variable is included in the prediction model or that the data selection was based on the variable. See Supplementary material B for background information on these graphs. 1A. Confounding: The confounding effect of previous hypertension (not included in the prediction model) induces a non-causal association between 'blood pressure medication' and 'cardiovascular disease'. 1B. Selection: By selecting only patients with a heart condition a non-causal association is induced between 'BMI' and 'unmeasured risk factors' and therefore with 'mortality'. 1C. mediation: By including a consequence of 'smoking' in the model (the mediator 'respiratory diseases'), part of the effect of 'smoking' on 'covid-19 related death' could be masked.*

| Defining the prediction estimand | | |
|---|---|---|
| **Estimand element** | **What does it mean?** | **Recommendations** |
| Target population | Individuals to whom the prediction model will be applied | Describe the patient group for whom all treatment options are possible and for whom a treatment decision is necessary. |
| Time point of intended use | The moment when the prediction model will be applied | Align time point of intended use with the moment of decision making. |
| Outcome and prediction horizon | The outcome of interest and the time horizon over which predictions are made | Choose an outcome and time horizon that is relevant to the treatment decision. |
| Predictors | The clinical and/or demographic patient factors to include in the prediction algorithm | Include predictors that are (readily) available at the time of treatment decision making. Ensuring that predictors are measured pre-treatment avoids including mediators as predictors (e.g., 'smoking and risk of covid-19 death'). Choose predictors separately from variables needed to adjust for confounding; these don't need to be the same. |
| Treatment option(s) | The target treatment options under consideration for decision making, in particular the option(s) under which estimates of potential risks are desired | Distinguish between baseline treatments and treatment options that represent longer exposure periods post-baseline. |
| **Assessing the data source** | | |
| **Design aspect** | **What does it mean?** | **Recommendations** |
| Sample selection | How individuals were included in the dataset | Selection bias may arise if inclusion in the data is (either directly or indirectly) related to the treatment and the outcome (e.g., 'BMI and mortality risk'). |
| Treatment assignment | The policy by which the target treatment was assigned in the data | Specify whether treatment was randomized or chosen by patients/doctors. How was adherence to the treatment? If the indication for treatment assignment is linked to the outcome then confounding is likely (e.g., 'blood pressure-lowering medication and cardiovascular risk'). |
| **Development and evaluation of the prediction algorithm** | | |
| **Analysis step** | **What does it mean?** | **Recommendations** |
| Assumptions | The assumptions required for valid prediction under interventions | Explicitly state and justify the assumptions that allow a ‘what if’ interpretation of predictions, e.g., no unmeasured confounding. |
| Handling of target treatment | How the treatment was handled in the analysis | Decide whether to use the treatment as a variable in the model, restrict on it (for baseline treatments only), censor or otherwise. Distinguish prevalent from incident use of treatment. |
| Handling of confounding | How confounding was addressed | Use causal inference methods to adjust for baseline and time-varying confounding when treatment was not randomized. |
| Time zero | Definition of baseline | Use the time of decision making as time zero and align eligibility, treatment assignment and start of follow at this time point to avoid time-related bias. |

*Table 1 Recommendations for prediction under interventions*

# Supplementary materials

## Contents

# Supplementary material A: Additional selection examples

Below we present two additional examples of how the interpretation of prediction models may depend on how the observational development data was selected. Just like in the BMI example presented in the main text, the bias in these examples is inflicted by selecting the data based on a third variable that has a causal relationship with both the treatment/exposure and the outcome. In the first example, below, the selection is directly influenced by treatment and outcome. In the second example, factors that influence sample selection influence the treatment and outcome.

## Exercise and covid-19 infection risk

During the covid-19 pandemic several prediction models were developed using data collected with mobile phone apps, to which participants could voluntarily contribute daily information on their activities and on covid-19 infection status (e.g., [1]). Suppose that one of the variables collected and used as predictor in a model for covid-19 infection was exercise level and that investigators would like to use this information to estimate potential risks if individuals were or were not to pursue a certain exercise level. It can be expected that app users are not a random sample from the population.[2] Suppose that health conscious people who exercise more are more likely to be contribute data to this app, and that at the same time people with symptoms of underlying covid-19 infection or who have recently been in contact with infected individuals were more inclined to use the app. This self-selection leads to artificially low estimated infection risks for individuals who exercise more, purely due to the way the data were collected. Therefore these risks are not useful to inform lifestyle changes in new individuals.

## Lifestyle and cancer risk in UK Biobank

McCarthy et al (2000) used the UK Biobank data to develop a prediction model for head and neck cancer.[3] Several unhealthy lifestyle factors were included as predictors in their model and showed to be associated with higher cancer risk. In their discussion they recommend that their personalized risk estimates 'could support discussions between dental professionals and patients regarding risk behaviours, such as smoking and alcohol consumption. This would also provide the opportunity to discuss health promoting behaviours, such as eating fresh fruit and vegetables and taking regular exercise.' This assumes a 'what-if' interpretation of the prediction model in which continuation of poor lifestyle behaviour leads to increased risk, and that this can be used to motivate lifestyle changes. However, lower estimated cancer risks for patients with healthier lifestyle could be in part due to selection mechanisms.[4] The UK Biobank enrolled over 500,000 individuals in the United Kingdom, but this was only 5.5% of all invited individuals. Participation was voluntary, resulting in a study group that was not representative of the UK population (e.g., the proportion of current smokers was lower).[5,6] UK Biobank participants also have lower rates of mortality and cancer incidence, which could be related to differences in observed lifestyle factors, but could also be due to differences in other biological  factors not recorded.[7] It is for instance known that individuals with higher socio-economic status (SES) were more inclined to participate and it is possible that the same holds for persons with biological risk factors.[7] Since SES is related to lifestyle habits, and biological risk factors relate to cancer incidence, selection based on both will introduce a non-causal association between lifestyle habits and cancer incidence. The McCarthy model did include a predictor representing SES of the area that participants lived in (Townsend deprivation index), but this may not fully capture individual participant's SES.

## Supplementary material B: Detecting confounding, selection and mediation bias with Directed Acyclic Graphs

Directed Acyclic Graphs (DAGs) are a helpful tool for detecting biasing structures, also in prediction models. A DAG is a graphical description of causal relationships (arrows) between variables (nodes). An assumed direct causal relation between two variables is depicted by a directed (solid) arrow between their nodes. The absence of an arrow between two nodes means that there is assumed to be no direct causal relation between the two variables. Indirect causal relations between variables appear as directed paths: paths in which arrows are traversed in their forward direction. Other types of paths do not represent causal relations. The "acyclic" in DAG means that such graphs do not contain a directed path from a node back to itself.

One can read off from a DAG if there are non-causal (biasing) associations between the target treatment/exposure and the outcome, by tracing open or blocked paths between these two nodes, taking into account the role of the other variables, e.g. when used for adjustment or when involved in selection.[8] We illustrate this below for the examples provided in the paper and the additional examples in this supplement. These structures can be helpful as a reference for future prediction studies to assess if similar biases are present.

A black outline of a node indicates that the variable is included in the prediction model or that the data selection was based on the variable. In the DAGs we only show the arrows and nodes relevant to the target treatment about which a decision is to be made; we use a dotted line to depict the (potential causal and non-causal) associations between other predictors and the outcome.

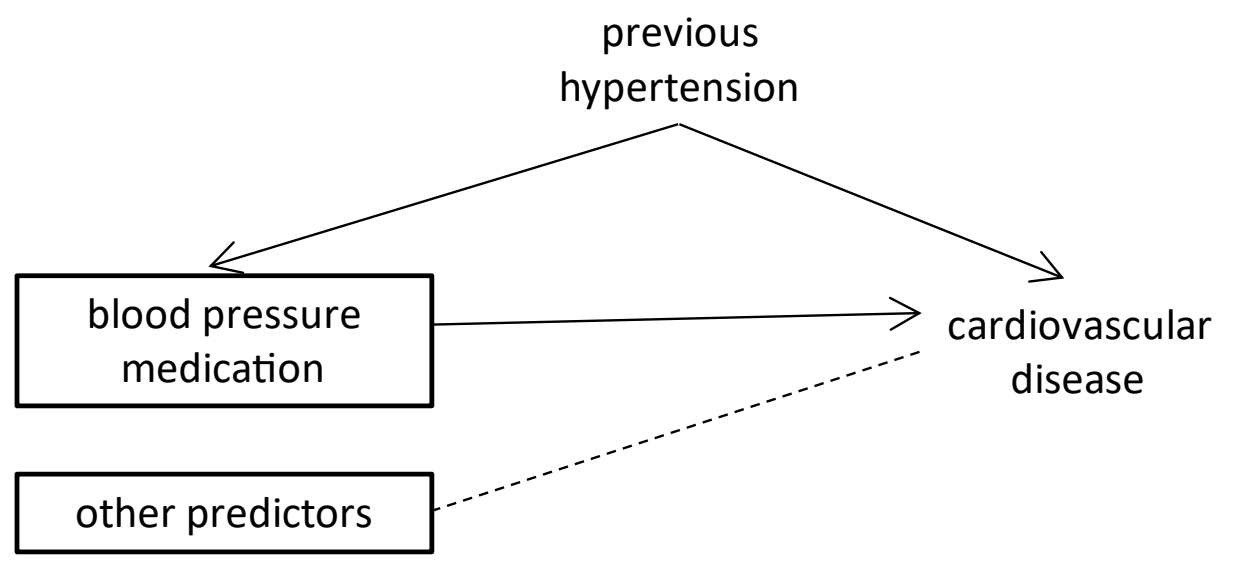

*Confounding –blood pressure lowering medication and cardiovascular risk*

The confounding effect of previous hypertension is reflected in the DAG by the non-causal path between 'blood pressure medication' and 'cardiovascular disease'. This path is induced by the confounder 'previous hypertension' that is not included in the prediction model.

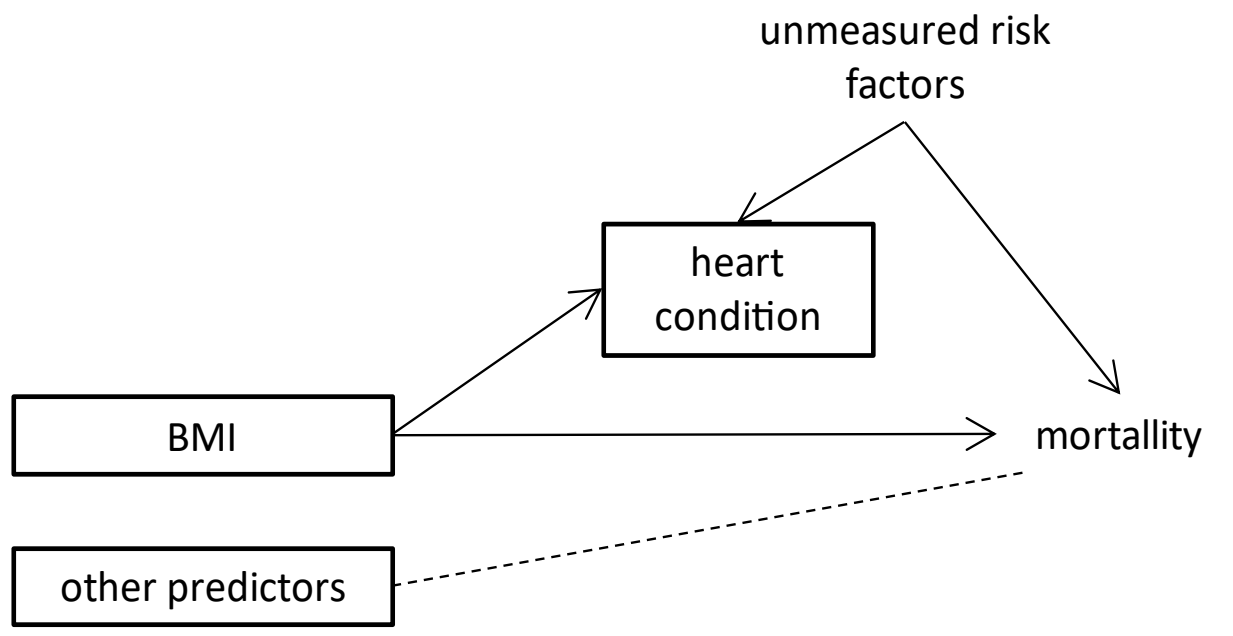

*Selection – BMI and mortality risk*

By selecting only patients with a heart condition (indicated by a box around it), a non-causal path is opened between 'BMI' and 'mortality'. The arrows from BMI and unmeasured risk factors meet (collide) at the selection node; the bias resulting from such sample selection is therefore also referred to as collider bias.

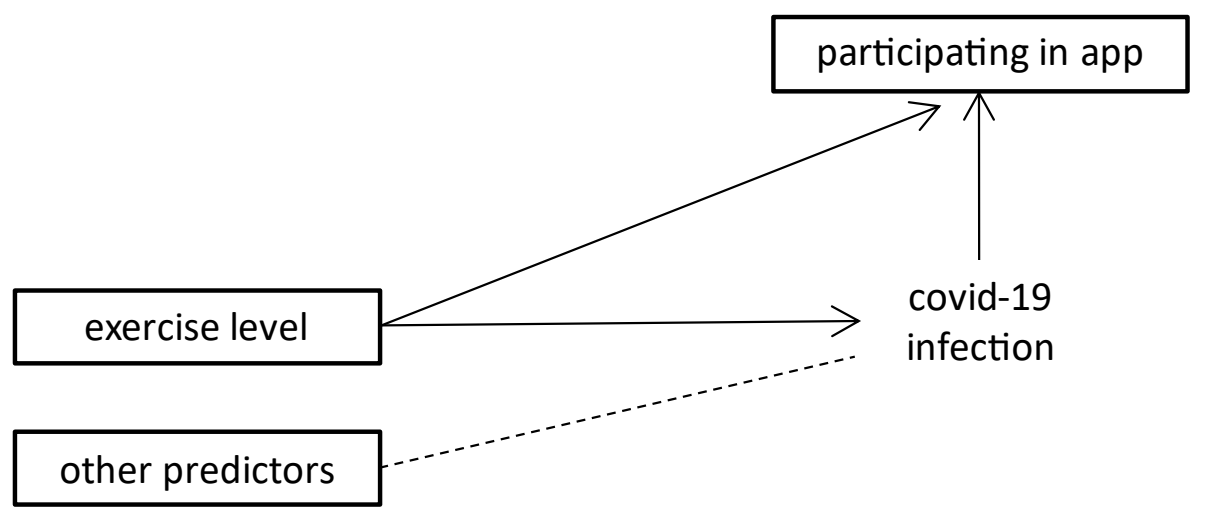

*Selection - exercise and covid-19 infection risk*

A non-causal association between 'exercise level' and 'covid-19 infection' is induced by selecting the data through voluntary participation in the app (again, a collider node).

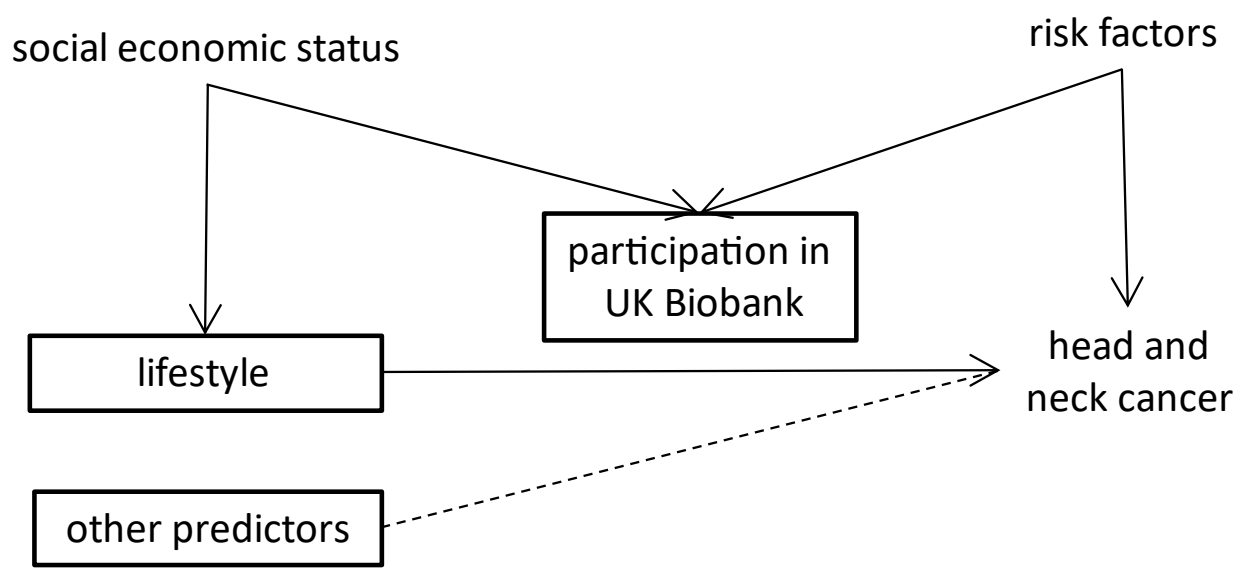

*Selection – lifestyle and cancer risk*

A non-causal association between 'lifestyle habits' and 'head and neck cancer' is induced by selecting the data through voluntary participation in UK biobank. This type of collider bias is also referred to as M-bias, due to the shape of the DAG.

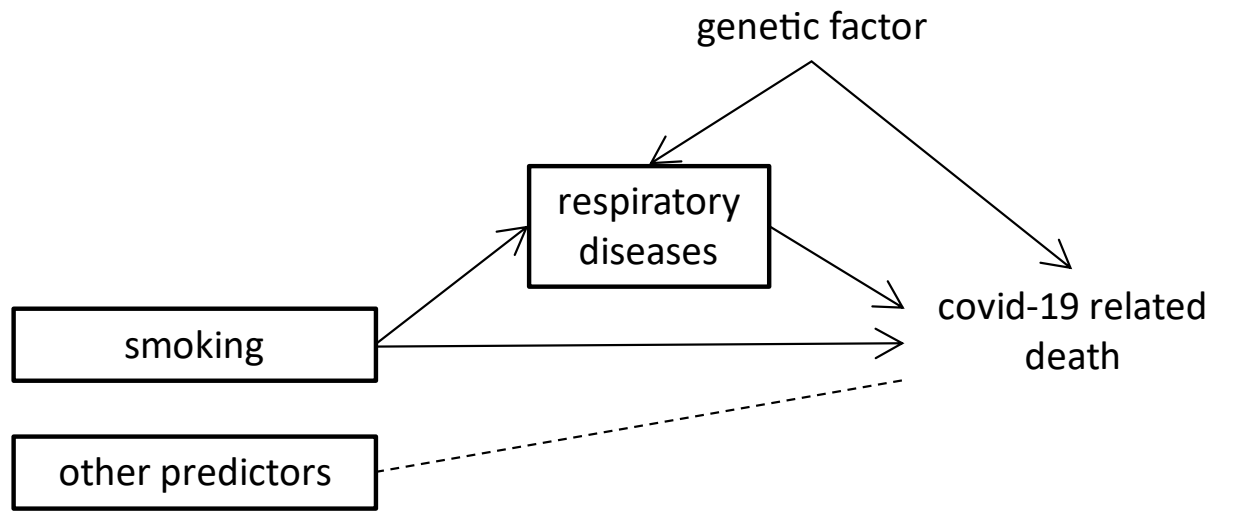

*Mediation – smoking and covid-19 death*

By including a consequence of 'smoking' in the model (the mediator 'respiratory diseases'), part of the effect of 'smoking' on 'covid-19 related death' is blocked. In addition, including respiratory diseases in the model may open a non-causal path through an unobserved 'genetic factor', introducing collider bias similar to the BMI example.

## References Supplementary Materials

1. Menni C, Valdes AM, Freidin MB, et al. Real-time tracking of self-reported symptoms to predict potential COVID-19. Nat Med. 2020;26:1037–40.

2. Griffith GJ, Morris TT, Tudball MJ, et al. Collider bias undermines our understanding of COVID-19 disease risk and severity. Nat Commun 2020;11:5749. doi:10.1038/s41467-020-19478-2

3. McCarthy CE, Bonnet LJ, Marcus MW, et al. Development and validation of a multivariable risk prediction model for head and neck cancer using the UK Biobank. International Journal of Oncology. 2020;57:1192–202.

4. Munafò MR, Tilling K, Taylor AE, et al. Collider scope: when selection bias can substantially influence observed associations. International Journal of Epidemiology. 2018;47:226–35.

5. Keyes KM, Westreich D. UK Biobank, big data, and the consequences of non-representativeness. Lancet (London, England). 2019;393:1297.

6. Stamatakis E, Owen KB, Shepherd L, et al. Is Cohort Representativeness Passé? Poststratified Associations of Lifestyle Risk Factors with Mortality in the UK Biobank. Epidemiology. 2021;32:179–88.

7. Fry A, Littlejohns TJ, Sudlow C, et al. Comparison of Sociodemographic and Health-Related Characteristics of UK Biobank Participants With Those of the General Population. American Journal of Epidemiology. 2017;186:1026–34.

8. Greenland S, Pearl J, Robins JM. Causal diagrams for epidemiologic research. Epidemiology. 1999;10:37–48.